 \documentclass[preprint2]{emuapj_aastex2}

 \usepackage{apjfonts}

\newcommand{\nevi}{\ion{Ne}{6}}

\newcommand{\mgvi}{\ion{Mg}{6}}

\newcommand{\ratio}{\ensuremath{f(\mathrm{Ne^{+5}})/f(\mathrm{Mg^{+5}})}}

\newcounter{hours}\newcounter{minutes}
\newcommand{\printtime}{%
  \setcounter{hours}{\time/60}%
  \setcounter{minutes}{\time-\value{hours}*60}%
  \thehours : \kern-0.4em \theminutes}

\newcommand\plotonefiddle[2]{%
 \centering
 \leavevmode
 \includegraphics[angle=#2,width={\columnwidth}]{#1}%
}%

\shortauthors{Edgar \& Esser}
\shorttitle{Non-equilibrium Ionization and FIP Effect Diagnostics}

\begin{document}


\title{Non-equilibrium Ionization and FIP Effect Diagnostics}


\author{
Richard J. Edgar and Ruth Esser
}
\affil{Harvard-Smithsonian Center for Astrophysics, Cambridge, MA 02138}



%%%%%%%%%%%%%%%%%%%%%%%%%%%%%%%%%%%%%%%%%%%%%%%%%%%%%%%%%%%%%%%%%%%%%%%%%
%%%%%%%%%%%%%%%%%%%%%%%%%%%%%%%%%%%%%%%%%%%%%%%%%%%%%%%%%%%%%%%%%%%%%%%%%
%%%%%%%%%%%%%%%%%%%%%%%%%%%%%%%%%%%%%%%%%%%%%%%%%%%%%%%%%%%%%%%%%%%%%%%%%
\begin{abstract}

% DRAFT \today.

We examine the accuracy of a common FIP effect diagnostic, the ratio
of \nevi\ to \mgvi\ lines in the solar transition region.  Since the two
ions have quite similar contribution functions near their maxima in
equilibrium, the ratio of these two ions is often taken to be
the abundance ratio of Ne and Mg.
First we compute non-equilibrium ionization effects on the ratio \ratio\ 
of ion fractions for a variety of simple flows through the transition region.
Calculating the spectral line ratios for a few examples, we then show
that non-equilibrium effects as well as temperature and density dependence
must be evaluated for each line ratio used in the diagnostics.

\end{abstract}




%%%%%%%%%%%%%%%%%%%%%%%%%%%%%%%%%%%%%%%%%%%%%%%%%%%%%%%%%%%%%%%%%%%%%%%%%
%%%%%%%%%%%%%%%%%%%%%%%%%%%%%%%%%%%%%%%%%%%%%%%%%%%%%%%%%%%%%%%%%%%%%%%%%
%%%%%%%%%%%%%%%%%%%%%%%%%%%%%%%%%%%%%%%%%%%%%%%%%%%%%%%%%%%%%%%%%%%%%%%%%
\section{Introduction}
   \label{sec:introduction}

\begin{figure}[!ht]
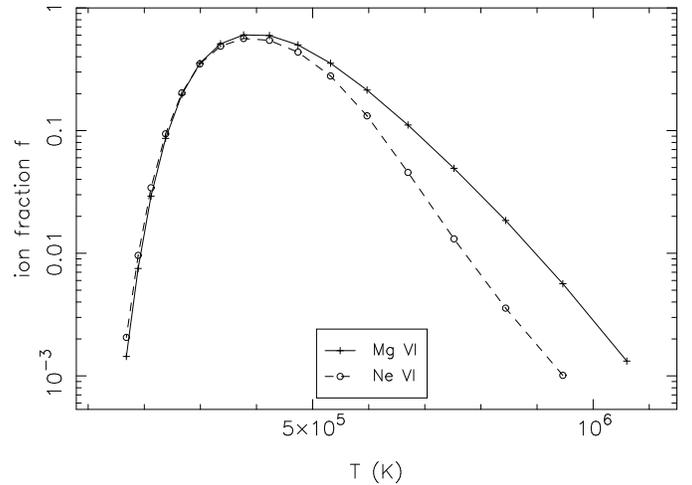

  %% \plotone{ratio.ps}
  \plotonefiddle{contrib.ps}{270}
  \caption[\nevi/\mgvi\ ratio]
          {Ion fractions for \nevi\ and \mgvi\ in coronal equilibrium.
          \label{fig:contrib}}
\end{figure}

In many regions of the solar atmosphere (including the transition
region, corona and solar wind) it appears that elements with a first ionization
potential (FIP) greater than 10 eV are enhanced relative to those with
$\mathrm{FIP} \la 10$~eV.
A common diagnostic  to quantify this effect is the measurement of the
ratios of emission lines of \mgvi\ (low FIP) and \nevi\ (high FIP)
(e.g. \citet{jordan98}; \citet{wf89}).  
% In equilibrium both ions have quite
% similar contribution functions (ion fractions vs temperature)
% over a wide temperature range (see Fig. \ref{fig:contrib}).
Both ions produce spectral lines in the neighborhood of
400 \AA\ (minimizing possible calibration problems), and have
very similar contribution functions in coronal equilibrium
(see Fig. \ref{fig:contrib}).  This has been characterized as one
of the best diagnostics for FIP-related abundances.

\begin{figure}[!ht]
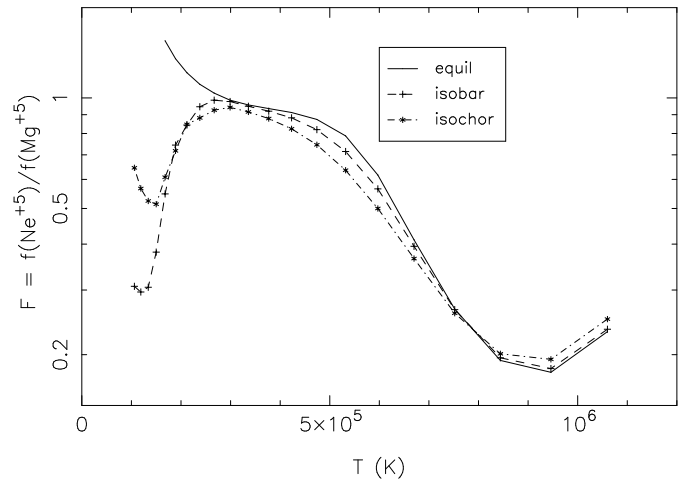

   %% \plotone{ratio.ps}
   \plotonefiddle{ratio.ps}{270}
   \caption[Ne VI/Mg VI ratio]
           {Ratio $F$ of ion fractions for coronal equilibrium,
 		and two cooling flow non-equilibrium ionization scenarios
 		(constant pressure and constant density).
           \label{fig:cooling}}
 \end{figure}

We calculate the ion fractions for both Ne and Mg ions, e.g.
$f(\mathrm{Ne^{+5}}) \equiv n(\mathrm{Ne^{+5}})/n(\mathrm{Ne})$
(Fig. \ref{fig:contrib}).
We then compute the ion ratio
$ F \equiv f(\mathrm{Ne^{+5}})/f(\mathrm{Mg^{+5}})$ in equilibrium
(Fig. \ref{fig:cooling}).
There is a wide plateau near $T=4\times 10^5$~K.  It is this feature which
is exploited when claiming this ion ratio should be insensitive to the
details of the structure of the corona.  
The density ratio depends on the elemental abundances
of Ne and Mg.  The photospheric
abundance of Ne is poorly known, and inferred from observations elsewhere
in the solar system.  However, the ratio of the total abundances of the 
two elements will alter
our computed ratio everywhere only by a constant factor.

It can be seen in Fig. 2 that the ratio of the two ions
at high temperatures is
very different from that at temperatures where the maximum in the ion
fraction curves occurs, even in the equilibrium situation. 
Before interpreting ratios of spectral lines formed by these ions as
abundance ratios it is, therefore, necessary
to establish that the plasma is predominantly at a temperature of
$\sim 4 \times 10^5$~K. To determine the distribution of ion fractions,
needed to interpret the line ratio,
one can observe a series of lines from multiple Ne and Mg ions
\citep{young97}.

In the present paper we presume that the elemental abundance ratio of Ne/Mg is
fixed, and examine the effect of non-equilibrium ionization (NEI) 
situations.
The code used to integrate the non-equilibrium ionization state of the
gas is described elsewhere \citep{edch86, gec88, eeb97}.  It uses
tabulated rate coefficients and cooling rates from
\citet{raymond77} as updated in 1993.  We compute the rates in
the low-density limit, which may affect the dielectronic recombination
rates.  The detailed shapes of the curves in Fig. \ref{fig:contrib}
depend on what set of ionization and recombination rate coefficients
are used, but the general conclusions presented here are independent
of such choices.

The relevant comparison is between the time (or distance) scale
over which the temperature changes significantly (such as the
temperature scale height) and the ionization or recombination
time (or distance) scale.  The distance scales are given by
\begin{equation}
    \lambda_{ion} = { u } / { n_e C_{ion}}, ~\mathrm{or}~
    \lambda_{rec} = { u } / { n_e \alpha_{rec}},
\end{equation}
where 
$n_e$ is the electron density, $u$ is the flow speed,
$C_{ion}$ and $\alpha_{rec}$ are ionization and recombination
rate coefficients in units of $\mathrm{cm^3~s^{-1}}$,
respectively (Maxwellian averaged $<\sigma v>$, where
$\sigma$ is the cross section and $v$ is the random
relative speed of electrons and ions).

\section{Non-Equilibrium Flows}
   \label{sec:extreme}

\subsection{Recombining}

As an example of a recombining
flow, we look at the situation described by \citet{edch86}, namely a
hot gas ($T_h \sim 10^6$~K) permitted to cool (and recombine) due
only to its own radiation.  For $T \la 10^6$~K, the gas cools faster
than it recombines, producing an over-ionized situation.  This is
similar to a microflare condition.

We have computed two cases, and the results are plotted
in Fig. \ref{fig:cooling}.  The two situations are cooling at constant
pressure (isobarically),
and at constant density (isochorically).  
Note that for $T \ga 2.5\times10^5$~K, while the details differ a bit,
the ion ratio $F$ is quite similar to that in equilibrium.  However, at
lower temperatures, the departure from equilibrium is quite
extreme, reaching perhaps an order of magnitude
in the isobaric case.

The time scale for temperature change is given by
the cooling time, $t_{cool} = T/(dT/dt)$.  Since both the recombination
and cooling rates vary directly with density, ``\textit{fluence}''
($\int n_e dt$, units $\mathrm{cm^{-3}~s}$),
is the natural independent variable instead of time.
(For example, for density $10^9~\mathrm{cm^{-3}}$, a
fluence of $10^{10}~\mathrm{cm^{-3}~s}$ is reached in 10~s.)
If the
recombination time $1/{n_e\alpha_{rec}}$ becomes longer than the
cooling time, or equivalently the recombination coefficient becomes
less than the inverse cooling fluence, 
$\alpha_{rec} \la 1/{n_e t_{cool}}$, recombination will lag behind
the temperature change.  This effect can be seen in Fig. \ref{fig:cooling}
for $T \la 2.5\times 10^5$~K.

\begin{figure}[!ht]
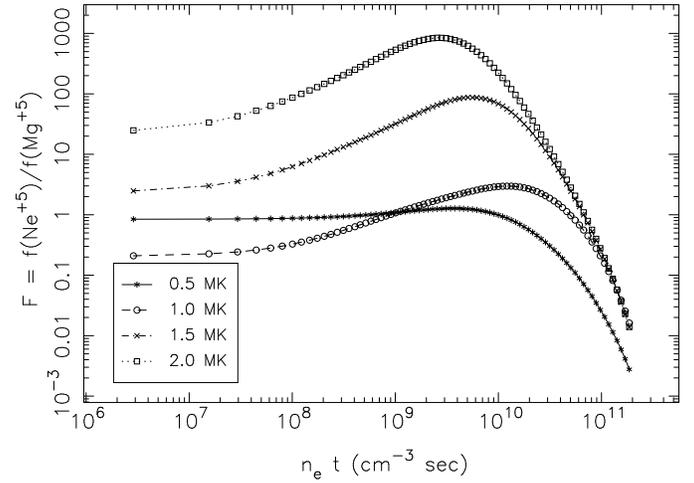

   %% \plotone{ratio.ps}
   \plotonefiddle{antishock.ps}{270}
   \caption[shock scenarios]
           {Ratio $F$ of ion fractions \textit{vs.}
 		fluence ($\int n_e~dt$) for four sudden cooling models.  
 		Initial high temperatures $T_h$ are given in MK.
           \label{fig:antishock}}
 \end{figure}

As a more extreme case of a recombining plasma, we consider
a gas taken to be at coronal equilibrium at an initial high temperature
$T_h\sim 10^6$~K, instantaneously reset to a cold temperature,
of $10^4$~K.  We then follow the recombination of Ne and Mg through the
various ionization states, as functions of the fluence $\int n_e~dt$.
We plot $F$ in Fig. \ref{fig:antishock}, and the
ionization fractions of $\mathrm{Ne^{+5}}$ and $\mathrm{Mg^{+5}}$ in Fig.
\ref{fig:nei_ionfrac} (downflow case).  Note that in the recombining case,
the curves for the two ions are
very different, which leads to a change in $F$ by an order of
magnitude during the recombination.  This is caused by the fact that
the Ne recombines through the +5 state sooner than the Mg does.
This example might represent an
extreme case of downflow through the transition region.

It is of course true that the details of this calculation depend on the
initial conditions, in particular on the ionization states of Ne and Mg
in the gas.  Real flows in the solar corona will be more complex, and
one must have an idea of the history of the gas before it begins the
downward flow through transition region temperatures in order to know
what initial conditions are appropriate.  However, the large
variability of $F$ in this example shows that downflows must be
treated with caution, when interpreting $F$ as an abundance
ratio.  (See also section \ref{sec:realistic}.)

\subsection{Ionizing}
   \label{sec:shock}

An extreme example of an ionizing plasma is
a shock, which instantaneously changes the temperature
from $10^4$~K 
to of order $10^6$~K.  The results are independent of the initial
temperature as long as it is low compared to the temperature range
at which the ions form (e.g. < 10$^5$ K, see Fig. 1). In the ensuing
flow, we watch as Ne and Mg atoms ionize up beyond the +5 state,
and compute the ion ratio.  

% We plot in Fig. \ref{fig:shock} the ion ratio $F$ for several different
% shock scenarios.  
% The ``post-shock'' temperature $T_h$ is 0.5, 1.0, 1.5, and 
% $2.0 \times 10^6$~K, and the ionization state of the gas
% is followed during the subsequent ionization up toward the new equilibrium.
% The figure clearly shows that the ion fractions vary by more than an order
% of magnitude through the zone where the ion fractions exceed $10^{-3}$,
% but the ion ratio is relatively flat and of order unity for $T_h=10^6$~K
% where the ion fractions
% have their maxima, as shown in Fig. \ref{fig:nei_ionfrac}.
% This figure shows the ion fractions for \nevi\ and
% \mgvi\ as functions of the fluence for the post-shock temperature  $T_h = 10^6$~K.
% For the upflow (ionizing case), the curves are quite similar.

\begin{figure}[!ht]
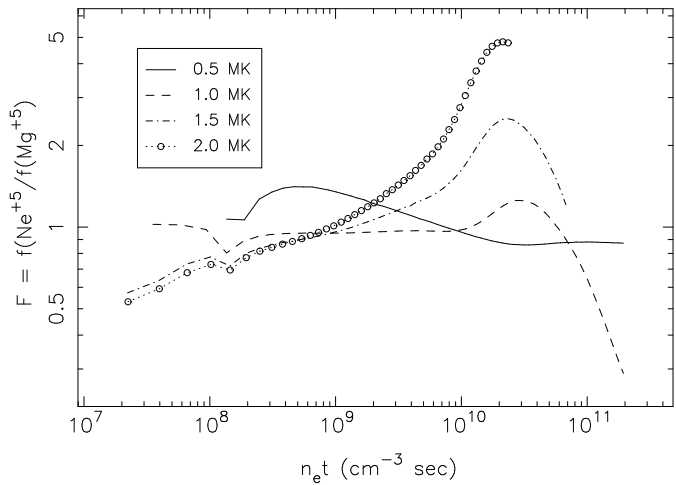

   %% \plotone{ratio.ps}
   \plotonefiddle{shock.ps}{270}
   \caption[shock scenarios]
           {The ratio $F$ = \ratio\ ion fractions \textit{vs.}
 		fluence ($\int n_e~dt$) for four shock models.  
 		Post-shock temperatures are given in MK.
           \label{fig:shock}}
 \end{figure}

\begin{figure}[!ht]
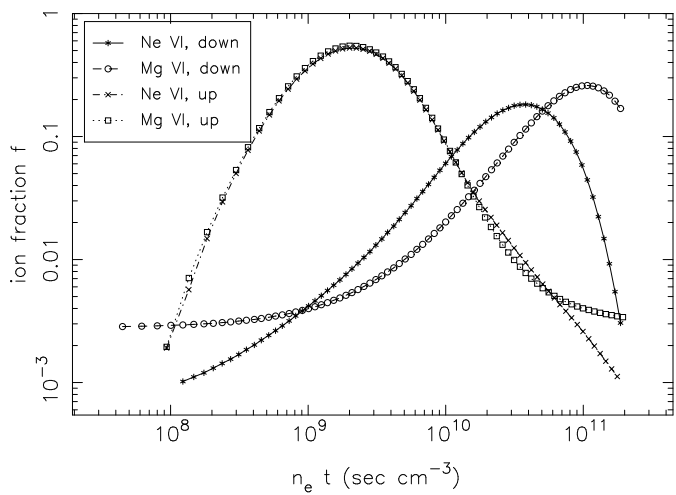

   %% \plotone{ratio.ps}
   \plotonefiddle{nei_ionfrac.ps}{270}
   \caption[NEI ion fractions]
           {Non-equilibrium ion fractions \textit{vs.}\/ fluence, for the case of
 	$T_h = 10^6$~K, for both ionizing (upflow) and recombining (downflow) models.
           \label{fig:nei_ionfrac}}
 \end{figure}

The ratio $F$ is shown as a function of fluence for four different shock
scenarios with ``post-shock'' temperatures $T_h$ of 0.5, 1.0, 1.5, and
$2.0 \times 10^6$~K in Fig. \ref{fig:shock}.
The gas is then followed towards the new equilibrium
situation.  For $T_h=2\times 10^6$~K, the ratio $F$ varies by about
one order of magnitude, but it is relatively constant and of order unity for
$T_h = 1.0\times 10^6$~K.  Ion fractions $f$ are shown for this case in
\ref{fig:nei_ionfrac}.


Upflows through the transition
region, modeled in the following section, are similar to this situation.

\subsection{Solar Wind Flows and Downflows}
   \label{sec:realistic}

\begin{figure}[!ht]
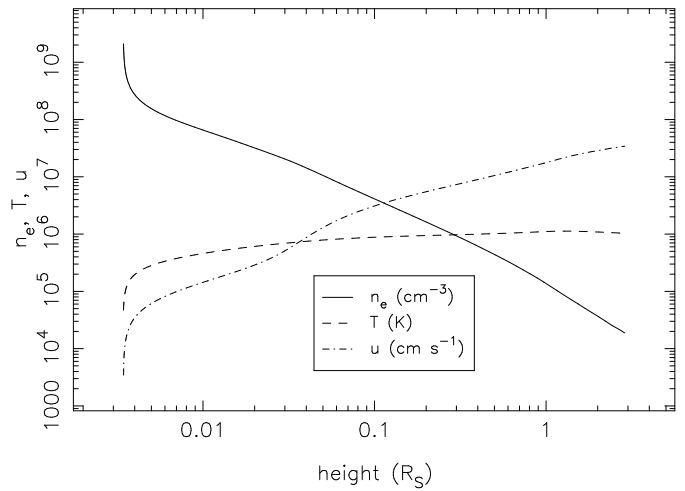

   %% \plotone{ratio.ps}
   \plotonefiddle{viggo_model.ps}{270}
   \caption[Viggo's model]
           {Density, temperature and velocity vs. height for the flow model
 	  of \cite{viggo97}.
           \label{fig:viggo_model}}
 \end{figure}

The characteristics of the solar wind flow in the critical
region, from the top of
the transition region at about $10^5$~K to the corona at about
$10^6$~K, are not well known. As an example we have chosen the electron
density, temperature and proton flow speed from a model by
\cite{viggo97} (Fig. \ref{fig:viggo_model}), as this model is the only
one to date that calculates the onset of the solar wind from the
mid-chromosphere to the corona self consistently. In the calculations
of the ion fractions shown in Figure \ref{fig:viggo_ratio}, we have
assumed that the Ne and Mg ions flow with the same speed as the
protons.
Note that this non-equilibrium case shows a
much wider region in temperature space where $F$ is of order
unity.  Essentially the same curve is obtained if we arbitrarily
multiply the flow speed by 10.  The equilibrium case (Fig. \ref{fig:cooling},
solid line) is shown as a comparison.

This scenario is somewhat similar to the shock flows presented
above (Sec \ref{sec:shock}), since
the flow is fast enough that the temperature changes significantly
in less than an ionization time:
\begin{equation}
   t_{ion} = {{1}\over{n_e C_{ion}}} \ga  {{1}\over{u}} {{T}\over{|\nabla T|}}.
\end{equation}
This similarity is perhaps not
surprising as the transition region is quite steep, and ends
at a temperature of the order of 10$^6$ K, as in one of the shock models (dashed line in Fig. 
\ref{fig:shock}). 

 \begin{figure}[!ht]
   \plotonefiddle{viggo_ratio.ps}{270}
   \caption[Viggo's ratios]
           {The ratio $F$ = \ratio\ ion fractions for the upflow model
 	  of \cite{viggo97}, and for the analogous time-reversed
 	  downflow model.
           \label{fig:viggo_ratio}}
 \end{figure}

% ALSO DESCRIBE THE DOWN FLOW VIGGO MODEL

Downflows are common in the transition region.  However, the details
such as velocity or temperature profile \textit{vs.} height are often
elusive.  Here we take as an example a time-reversed
version of the upflow model above.  The ionization
state of the gas at the beginning of the downflow is needed for the
model, but depends on the prior history of the gas.  We presume the
gas is in coronal equilibrium at two temperatures, 0.6 and
$0.8\times10^6$~K.  The resulting ion ratios $F$ are plotted in
Fig. \ref{fig:viggo_ratio}.  They start (on the right of the figure)
at the equilibrium curve (the assumed initial condition), and
move to the left toward lower temperatures with time.  Starting from
$0.8\times 10^6$~K, the ion ratio starts at low values and comes up
to nearly unity.  The details are somewhat different from the
sudden cooling model discussed above (Sec. \ref{sec:extreme})
due to the fact that the flow decelerates towards lower temperatures as the
density rises rapidly.  This causes the ionization state to approach
coronal equilibrium at the end of the flow.  Nonetheless, we
see that this downflow also has an ion ratio $F$ significantly
different from unity over the part of the flow where the ions are
abundant.

\subsection{Emissivity Calculations}
   \label{sec:emissivity}

In this section we discuss the intensity ratios of the spectral lines
most commonly used for abundance ratio determination for these two ions.
The \ion{Ne}{6} lines are transitions from
$\mathrm{2s2p^2~^2P_{3/2} \to 2s^22p~^2P_{3/2,1/2}}$
($\lambda\lambda 401.93,399.82$), and
$\mathrm{2s2p^2~^2P_{1/2} \to 2s^22p~^2P_{3/2,1/2}}$
($\lambda\lambda  403.26, 401.14$),
while the \ion{Mg}{6} lines connect
$\mathrm{2s2p^4~^4P_{1/2,3/2,5/2} \to 2s^2 2p^3~^4S_{3/2}}$
($\lambda\lambda 399.28, 400.66, 403.31$).
Emissivity calculations were performed using the APEC code \citep{smith00},
with atomic data taken from the CHIANTI database \citep{landi99}.
Energy levels for both \ion{Ne}{6} and \ion{Mg}{6} are taken from 
\cite{martin95}.
Collision and oscillator strengths for \ion{Ne}{6} are from
\cite{zhang94} and \cite{dankwort78}, respectively.
Those for \ion{Mg}{6} are taken from \cite{bhatia80}.

\begin{figure}[!ht]
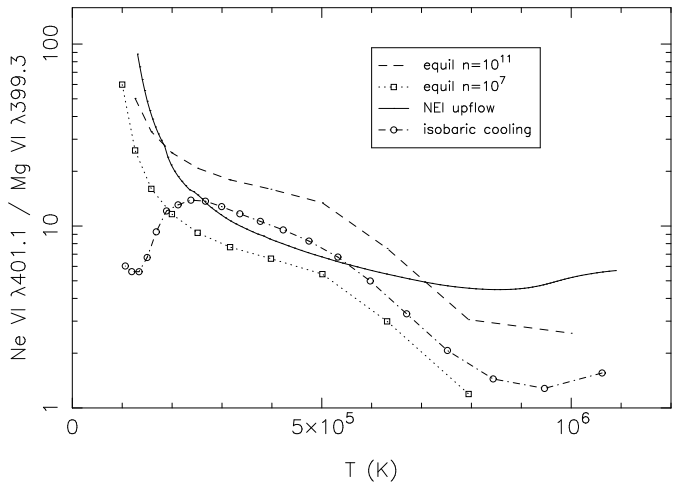

   %% \plotone{ratio.ps}
   \plotonefiddle{lineratio.ps}{270}
   \caption[Line ratios]
           {The ratio \ion{Ne}{6} $\lambda 401.1$/\ion{Mg}{6} $\lambda 399.3$
	 of line emissivities for coronal equilibrium at two densities,
	for the isobaric cooling case, and for the upflow model
 	  of \cite{viggo97}.
          \label{fig:line_ratio}}
 \end{figure}

The interpretation of emissivity ratios is even more complicated,
as the emissivities of these lines, and their ratios, depend on
both density and temperature. 
As examples, we plot in Fig. \ref{fig:line_ratio}, the ratio 
\ion{Ne}{6}~$\lambda 401.14$/\ion{Mg}{6}~$\lambda 399.28$
for three of the above models: equilibrium ionization, isobaric cooling
(with pressure $n_e T = 2.6\times10^{14} ~\mathrm{cm^{-3}~K}$),
and the \cite{viggo97} upflow model.  To demonstrate the density
dependence, the equilibrium curve is shown for two densities,
$10^7$ and $10^{11}~\mathrm{cm^{-3}}$.  Note that the region of weak
temperature dependence ($T \approx 2-5 \times 10^5$~K; \textit{e.g.}
Fig. \ref{fig:cooling}) has more or less disappeared.

These examples demonstrate that non-equilibrium effects as well as
temperature and density dependence must be evaluated for each line
ratio used in the diagnostic.

\section{Summary and Discussion}
   \label{sec:discussion}

The $\mathrm{Ne^{+5}}$ to $\mathrm{Mg^{+5}}$
line ratio diagnostic is generally based on the assumption that the
ions are in ionization equilibrium and that the plasma is predominantly
at temperatures between 2 and $5 \times 10^5$~K. Below and above that
range, temperature effects are important (e.g. Fig. \ref{fig:cooling}).
The goal of the
present paper was to investigate the effects of non-equilibrium
situations on $F \equiv f(\mathrm{Ne^{+5}})/f(\mathrm{Mg^{+5}})$.
If the plasma is
not in equilibrium, assumptions regarding the plasma dynamics and
prior history have to
be made, since plasma parameters describing these flow dynamics are often
unknown.  

To shed some light on NEI effects we have
considered a number of idealized cases. Some of these cases,
isobaric and isochoric cooling (Fig. \ref{fig:cooling}) and ionizing
shock situations (Fig. \ref{fig:shock}) maintain an ionization balance
for the two ions extremely close to the equilibrium values.  In the
presence of outflow, as in a solar wind situation where the flow speed
in the transition region is small $\lesssim 10~\mathrm{km~s^{-1}}$,
the ratio of the two
ions actually remains close to 1 over a larger temperature range (Fig.
\ref{fig:viggo_ratio}, solid line). 
However, calculating the spectral line ratios for a few examples
shows that this effect vanishes due to the presence of a density dependence
(Fig. \ref{fig:line_ratio}) which exists for both equilibrium and
NEI conditions.

In situations resembling downflows, on the other hand,
in which the plasma cools quickly (Figs. \ref{fig:antishock}
and \ref{fig:viggo_ratio}, dash and dash dotted lines), the
deviation from equilibrium might be quite significant. The two cases
shown in Fig. \ref{fig:viggo_ratio} demonstrate that in the downflowing
case, the ion ratio depends very much on the upper boundary condition which
is not known.  We have made the
simplifying assumption that the plasma is in equilibrium at the
temperature where it starts to downflow. This does not at all have to
be the case. The plasma could have been out of equilibrium in
the upflowing process, and since ionizing and recombining plasmas
behave differently, could depart from equilibrium even further. 

The calculations presented in this paper demonstrate that temperature
and density dependence as well as non-equilibrium effects must be
carefully evaluated for each line ratio used in the diagnostic.

%%%%%%%%%%%%%%%%%%%%%%%%%%%%%%

%% Included in this acknowledgments section are examples of the
%% AASTeX hypertext markup commands. Use \url without the optional [HREF]
%% argument when you want to print the url directly in the text. Otherwise,
%% use either \url or \anchor, with the HREF as the first argument and the
%% text to be printed in the second.

\acknowledgments

This work was supported by NASA contract NAS8-39073, and
NASA grant NAG5-7055.  We thank Randall Smith for the pre-release
use of the APEC code, and the anonymous referee for valuable
suggestions.

\end{document}